\documentclass[universe,article,accept,pdftex,moreauthors]{Definitions/mdpi} 
\firstpage{1} 
\pubvolume{1}
\issuenum{1}
\articlenumber{0}
\pubyear{2026}
\copyrightyear{2026}
\externaleditor{Firstname Lastname} % More than 1 editor, please add `` and '' before the last editor name
\datereceived{5 January 2026} 
\daterevised{22 January 2026} % Comment out if no revised date
\dateaccepted{ } 
\datepublished{ } 
\Title{SpecZoo: An AI-Powered Platform for Spectral Analysis and Visualization in Science and Education}

\Author{Yuanhao %MDPI: 1. Please carefully check the accuracy of names and affiliations. 2. The name of this author is different from the one submitted online at susy.mdpi.com. Please confirm which one is correct. Same for other highlighted names.
 Pu $^{1,2}$, Guohong 
 Lei $^{2,3}$, Yang Xu $^{4}$, Xunzhou 
 Chen $^{1,2}$ and Haijun 
 Tian $^{1,2,}$*}

\AuthorNames{Yuan-Hao Pu, Guo-Hong Lei, Yang Xu, Xun-Zhou Chen and Hai-Jun Tian}

\address{%
$^{1}$ \quad School of Science, Hangzhou Dianzi University, Hangzhou 310018, China; 231070086@hdu.edu.cn %MDPI: We added and highlighted the email addresses here according to those submitted online at susy.mdpi.com. Please confirm.(right)
 (Y.-H.P.); cxz@hdu.edu.cn 
 (X.-Z.C.)\\
$^{2}$ \quad Zhejiang Branch of National Astronomical Data Center, %MDPI: Please check that the address information is complete. The provided information should be arranged from subordinate to superior.（right)
 Hangzhou 310018, China; \linebreak  leiguohong@ctgu.edu.cn\\
$^{3}$ \quad School of Ethnic Studies, China Three Gorges University, %MDPI: For universities, the department/school/faculty/campus is required. Please try to provide this information.(College information of the university has been added)
 Yichang 443002, China\\
$^{4}$ \quad National Astronomical Observatory ,  %MDPI: it is recommended to replace this with a comma.
 Chinese Academy of Sciences, Beijing 100012, China; yxu@nao.cas.cn 
}

\corres{Correspondence: hjtian@hdu.edu.cn}

\abstract{Astronomical spectra, which encode rich astrophysical and chemical information, are fundamental to understanding celestial objects and universal laws. The advent of large-scale spectroscopic surveys, generating tens of millions of spectra, presents significant challenges for efficient data processing and analysis. To address these challenges, we develop an AI-powered platform (named ``SpecZoo'') for spectral visualization and analysis. This platform integrates modern information technology and machine learning to lower the barrier to spectral data utilization and enhance research efficiency. Its core functionalities include interactive visualization, automated spectral classification, physical parameter measurement, spectral annotation, and multi-band/multi-modal data fusion, all supported by flexible user and data management systems. It has become an essential tool for the National Astronomical Data Center, directly supporting spectral data processing and research for major projects including LAMOST, SDSS, DESI, and so on. Furthermore, the platform demonstrates strong potential for science-education integration, providing a novel resource for cultivating talent in astronomy and data science.}


\begin{document}

%%%%%%%%%%%%%%%%%%%%%%%%%%%%%%%%%%%%%%%%%%
\section{Introduction} \label{sec:intro}
Spectroscopic observations provide a fundamental means of probing the physical conditions of celestial objects. The detailed shapes and strengths of spectral features allow precise measurements of chemical abundances, effective temperatures, surface gravities, stellar ages,  kinematics, and so on. These parameters are crucial for understanding the formation and evolution of stars and stellar populations, constraining the structure of the Milky Way, and tracing the assembly histories of galaxies across cosmic time.

Modern large-scale spectroscopic surveys have transformed observational astrophysics by delivering unprecedented volumes of high-quality spectral data. Examples include the Large Sky Area Multi-Object Fiber Spectroscopic Telescope (LAMOST;~\cite{zhao2012lamost}), the Sloan Digital Sky Survey (SDSS;~\cite{kollmeier2019sdss}), the Dark Energy Spectroscopic Instrument (DESI;~\cite{levi2019dark}) {and the Global Astrometric Interferometer for Astrophysics} (Gaia;~\cite{lindegren1996gaia}). SDSS, through successive generations and most recently SDSS-V (2020--2025), is obtaining optical and near-infrared spectra for more than six million objects, enabling detailed investigations of Galactic structure, stellar populations, exoplanets, compact objects, and even the Universe. Complementing these, Gaia provides astrometric, photometric, and spectroscopic data for over 1.8 billion sources, delivering precise radial velocities and stellar parameters that are essential for studying the kinematics, chemistry, and formation history of the Milky Way. LAMOST, China’s first major national spectroscopic facility, combines a large aperture with a wide field of view and has released over twenty million spectra, providing an unparalleled resource for Milky Way archaeology and stellar parameter inference. DESI, operational since 2021, deploys 5000 robotically positioned fibers and collects over 100,000 spectra per night; its first-year data yielded a three-dimensional map containing 18.7 million galaxies, quasars, and stars, offering a transformative dataset for cosmology. Despite these advances, the vast and complex data produced by modern surveys pose new challenges for storage, visualization, and analysis.

To date, there remains no comprehensive international platform for spectroscopic data visualization, analysis, and management that simultaneously meets the needs of astronomical research, education, and public outreach. Many early-generation software packages---such as VOSpec~\cite{laruelo2008vospec}, SpecView~\cite{busko2000specview}, SPLAT~\cite{vskoda2014spectroscopic}, and CASSIS~\cite{lebouteiller2011cassis}---were developed on Java-based architectures. These tools suffer from cumbersome deployment, strong dependence on local computing environments, and limited functionality; moreover, they have not evolved into integrated research–education platforms, leading to their limited adoption in modern astronomical workflows. {While SkyPortal}~\cite{van2019skyportal} {excels in managing time-domain targets and collaborative workflows, they share a common limitation: as general-purpose observation management frameworks, their core designs are not optimized for the deep interactive analysis of spectroscopic data or research-grade parameter extraction}. They lack built-in, advanced AI-powered capabilities dedicated to spectral features, such as automated classification, parameter estimation, and anomaly detection. Furthermore, their collaboration models are generally confined to task assignment and discussion within project teams, failing to establish an open ecosystem for spectroscopic analysis that seamlessly integrates professional research with public participation.

{In comparison to general-purpose citizen science platforms such as Zooniverse}~\cite{simpson2014zooniverse}, {existing spectral tools lack sufficient depth in specialization and interactive analysis. While Zooniverse has successfully enabled multi-project management and public participation, its design is not optimized for complex spectral data analysis and research-level parameter measurement. Our developed platform, SpecZoo, aims to bridge this gap by providing a specialized web-based platform focused on the domain of spectroscopy, integrating artificial intelligence with collaborative functionalities.}

In recent years, machine learning methods have demonstrated significant advantages in astrophysical spectral classification, parameter estimation, and the discovery of rare celestial objects, leading to the proliferation of algorithms and models based on techniques such as deep learning and transfer learning. However, these studies have largely focused on the algorithms themselves and have not been adequately integrated into user-friendly, open, and collaborative analysis platforms, resulting in barriers to their application within practical scientific workflows.

To address these challenges, we develop an AI-powered, web-based platform (named ``SpecZoo'') for spectral data visualization and analysis. Built on a modern front-end/back-end architecture, the platform provides full functionality through a standard web browser, eliminating technical barriers and simplifying access. SpecZoo integrates a suite of AI algorithms for automated spectral classification, parameter estimation, redshift measurement, and anomaly detection and supports advanced capabilities such as multi-modal data fusion, interactive visualization, and collaborative annotation workflows. In scientific applications, SpecZoo enables targeted object searches, sample construction, and efficient validation of rare or peculiar celestial objects. For educational use, the platform facilitates research-oriented teaching, allowing students to explore authentic spectral datasets and perform detailed data annotation.

As the foundational article of the SpecZoo series, this work describes the platform’s architectural design, core functionalities, and operational capabilities. Building upon this foundation, subsequent studies will leverage machine‑learning techniques and the SpecZoo platform to systematically search for or validate rare celestial objects. These research activities will be integrated into research‑oriented teaching, with the ultimate goal of establishing SpecZoo as a comprehensive spectral “zoo” for both scientific and educational communities.

The remainder of this paper is organized as follows: Section~\ref{sec:style} describes the design philosophy and modular architecture of SpecZoo, outlining its foundational pillars of scientific empowerment, educational innovation, and sustainable data ecosystem development; Section~\ref{sec:floats} details the platform's core functionalities and technological advancements, covering basic role and data management, spectral template libraries, and advanced AI-powered features; Section~\ref{sec:scientific} illustrates several scientific use cases; and Section~\ref{sec:education} discusses the platform's educational applications, focusing on its integration into structured spectral identification training and research-oriented teaching modules that cultivate practical data analysis skills. Finally, we give the conclusion in \linebreak  Section \ref{sec:conclusion}.

\section{Design of SpecZoo} \label{sec:style}

SpecZoo is an AI-powered spectral platform that enhances previous system architectures~\cite{Lei2018-ExpertSpectralPlatform} by integrating large-scale spectral data analysis with research and educational applications. In this section, we describe the platform's design philosophy, modular architecture, and technical details, highlighting how SpecZoo addresses challenges in managing, visualizing, and analyzing massive spectroscopic datasets.

\subsection{Design Philosophy}

\begin{figure}[H]
\includegraphics[width=0.95\linewidth]{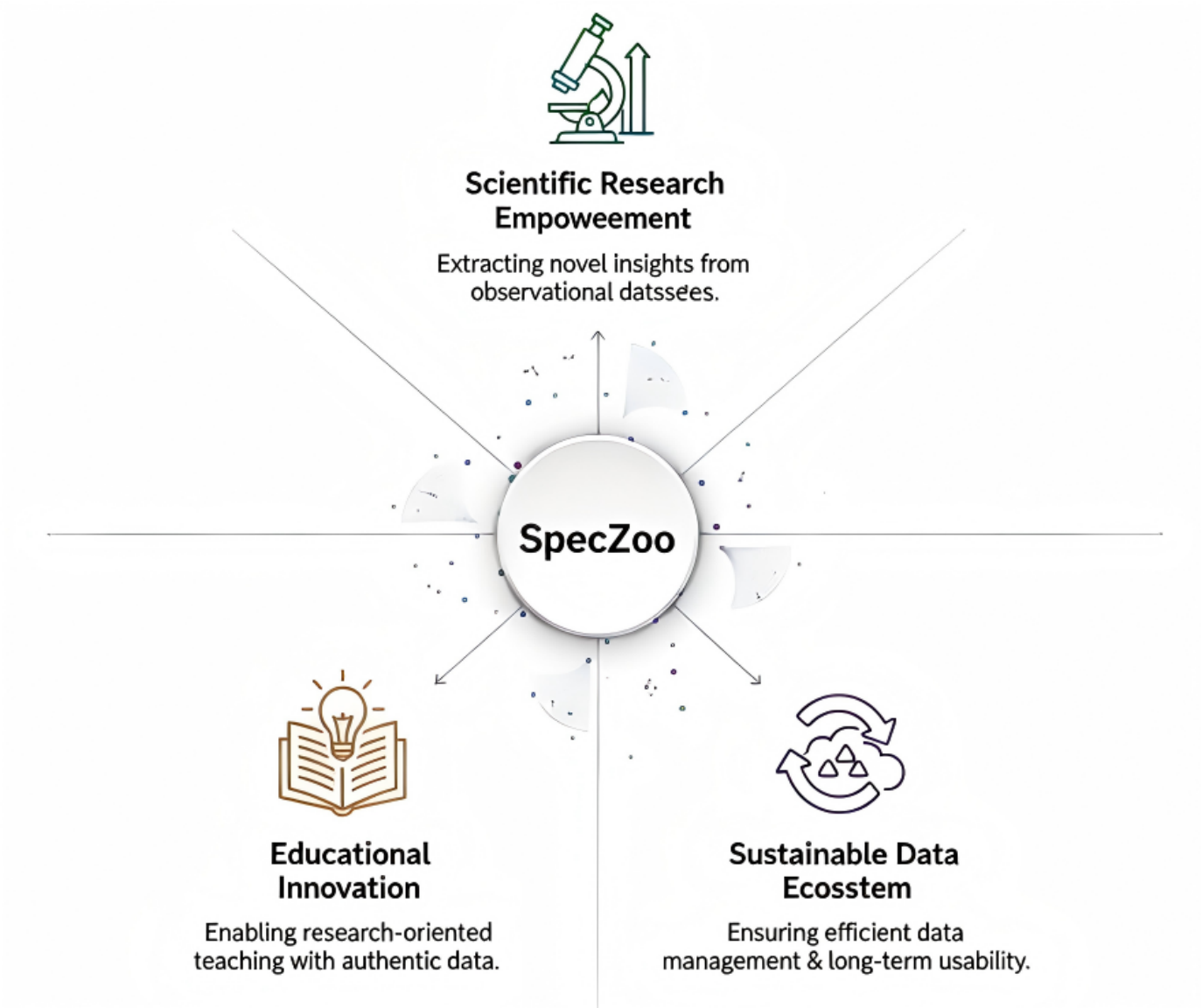}
\caption{SpecZoo %MDPI: Please check if dots, colors, arrows or colors need to be explained in caption.(It's not necessary; they just want it for aesthetic purposes.)
 design philosophy. The platform integrates AI-powered analysis, educational innovation, and sustainable data management to address the challenges posed by large-scale spectral data.\label{fig:2}}
\end{figure}

SpecZoo is built on three foundational pillars: scientific research empowerment, educational innovation, and sustainable data ecosystem development. These principles guide both the architectural and functional design of the platform. By emphasizing scientific research, SpecZoo enables users to extract novel insights from each observational dataset. At the core of the SpecZoo design, illustrated in Figure~\ref{fig:2}, are two foundational pillars supporting a sustainable research-teaching ecosystem. First, the sustainable data pillar establishes a robust infrastructure for efficient data management, sharing, and long-term preservation. Centered on these foundations is the educational pillar, which facilitates research-oriented teaching by enabling students to work directly with authentic spectral data, thereby cultivating essential \linebreak  analytical skills.

\subsection{Modular Architecture of SpecZoo}

SpecZoo adopts a modular, front-end/back-end separation architecture, consisting of four core layers: the User Roles Layer, the Visualization and Label Layer, the Data Node Layer, and the AI Layer, as shown in Figure~\ref{fig:3}.

The {User Roles Layer%MDPI: Please confirm if the bold is unnecessary and can be removed. The following highlights are the same.
 manages user authentication, permissions, task assignments, annotation review, and statistical reporting. Single sign-on via the OAuth protocol~\cite{hardt2012oauth} allows registered users of the National Astronomical Data Center (NADC) to access the platform without additional registration steps.

The {{Visualization and Label Layer} 
 provides interactive online spectral visualization and analysis. It supports spectral map display, spectral line annotation, multi-band data fusion, AI-assisted classification recommendations, template matching, and redshift or velocity measurements. To enhance data quality, the layer incorporates wavelet-based noise reduction and {a tool for removal of 3$\sigma$ sky lines}. Visualization interfaces are implemented with {ECharts}~\cite{li2018echarts}, enabling multi-terminal interaction.

The {{Data Node Layer} handles storage and access control for user-submitted spectral data and star catalogs. Data visibility is managed through three modes: PublicDB (public database), GroupDB (group-shared database), and MyDB (user-private database). {Large files} are uploaded using segmentation and breakpoint-resume mechanisms, ensuring stability and efficiency.

The {{AI Layer} 
 integrates machine learning methods for tasks including spectral classification, stellar feature extraction, calculation of stellar atmospheric parameters, and redshift measurements for galaxies and quasars. Computationally intensive operations run on a dedicated server, with results transmitted to the front-end for visualization and further analysis.

\begin{figure}[H]
\includegraphics[width=\linewidth]{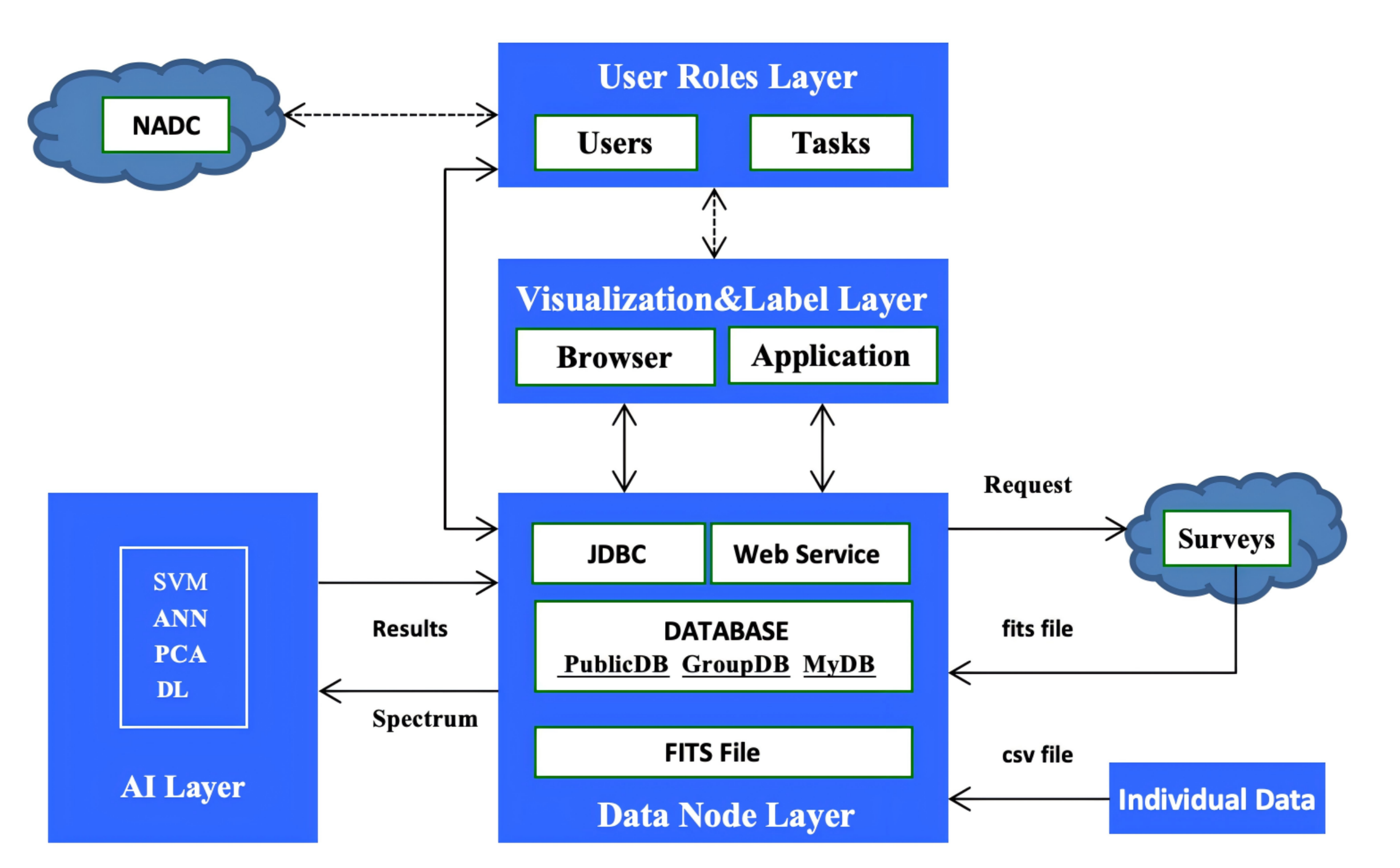}
\caption{{Architectural} %MDPI: Please check if background colors need extra explanation.(not neccessary)
 design of SpecZoo. The platform consists of four layers: User Roles, Visualization and Label, Data Node, and AI, enabling interactive analysis, data management, and AI-powered processing.\label{fig:3}}
\end{figure}

The front-end is implemented using the Vue.js framework~\cite{hanchett2018vue}, chosen for its interactivity and maintainability. The back-end employs SpringBoot~\cite{boaglio2017spring}, simplifying configuration management and dependency handling. Data persistence is handled by MyBatis~\cite{ho2012using}, which maps Java objects to MySQL database records {(SpecZoo emphasizes stable data acquisition and management, visualization, long-term maintainability, and scalability in its technical design. Vue.js was selected for its low learning curve and well-structured component architecture, facilitating efficient development of complex data interfaces. Spring Boot streamlines backend implementation with its convention-over-configuration approach and robust ecosystem. MySQL provides a reliable and cost-efficient solution for structured metadata storage. MyBatis is employed to maintain explicit SQL control, which is crucial for performance-sensitive astronomical queries and complex data operations. Collectively, these established technologies constitute a sustainable and extensible stack that aligns with the collaborative and evolving requirements of scientific research)}. This architecture ensures high performance, scalability, and system stability. SpecZoo is currently operational at the {NADC}%MDPI: 1. We revised footnotes to endnotes, please check and confirm. 2. Please add the access date (format: Date Month Year), e.g., accessed on 1 January 2020.
\endnote{\url{https://nadc.china-vo.org/speczoo-system/}}.

\section{Implementation of SpecZoo} \label{sec:floats}

\subsection{Basic Functions}

{{Role Management.} 
 The Workgroup function serves as the cornerstone for collaborative work within SpecZoo. SpecZoo employs a hierarchical user-role system with four levels: administrators, group leaders, group managers, and regular users. Group leaders form teams and define spectroscopic tasks, group managers assist with task assignment and monitoring, and regular users have the flexibility to join multiple workgroups, where they can share data and engage in cooperation, thereby fostering a dynamic scientific community. This structure ensures organized collaboration and efficient task management. Specific permissions are as shown in Figure~\ref{quanxian}.

\begin{figure}[H]
\includegraphics[width=0.9\linewidth]{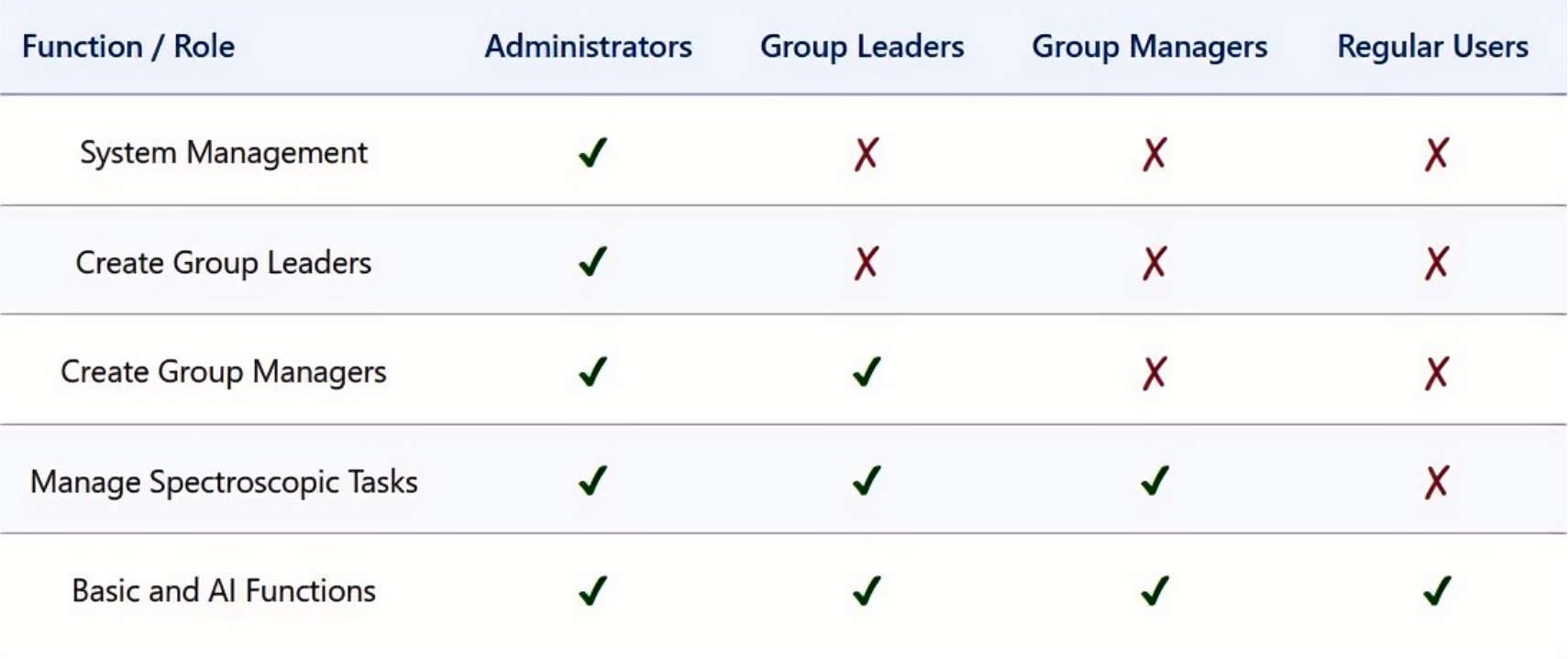}
\caption{{Specific} %MDPI: Please check if background color need extra explanation(not necessary)
 permissions of characters.\label{quanxian}}
\end{figure}

{{Spectral Templates and Characteristic Lines.} 
 SpecZoo provides predefined spectral templates for stars, galaxies, quasars, and other common astronomical objects, alongside a library of characteristic spectral lines. As an example, a quasar template is illustrated in Figure~\ref{fig:template}. SpecZoo currently hosts a collection of 334 standard templates. These resources can be flexibly updated by administrators via the backend and are readily accessible to users, who may actively employ them for visual searches of specific or rare celestial objects. They serve to standardize spectral classification and line identification, thereby improving analytical consistency and lowering the entry barrier for less experienced users.

\begin{figure}[H]
\includegraphics[width=0.9\linewidth]{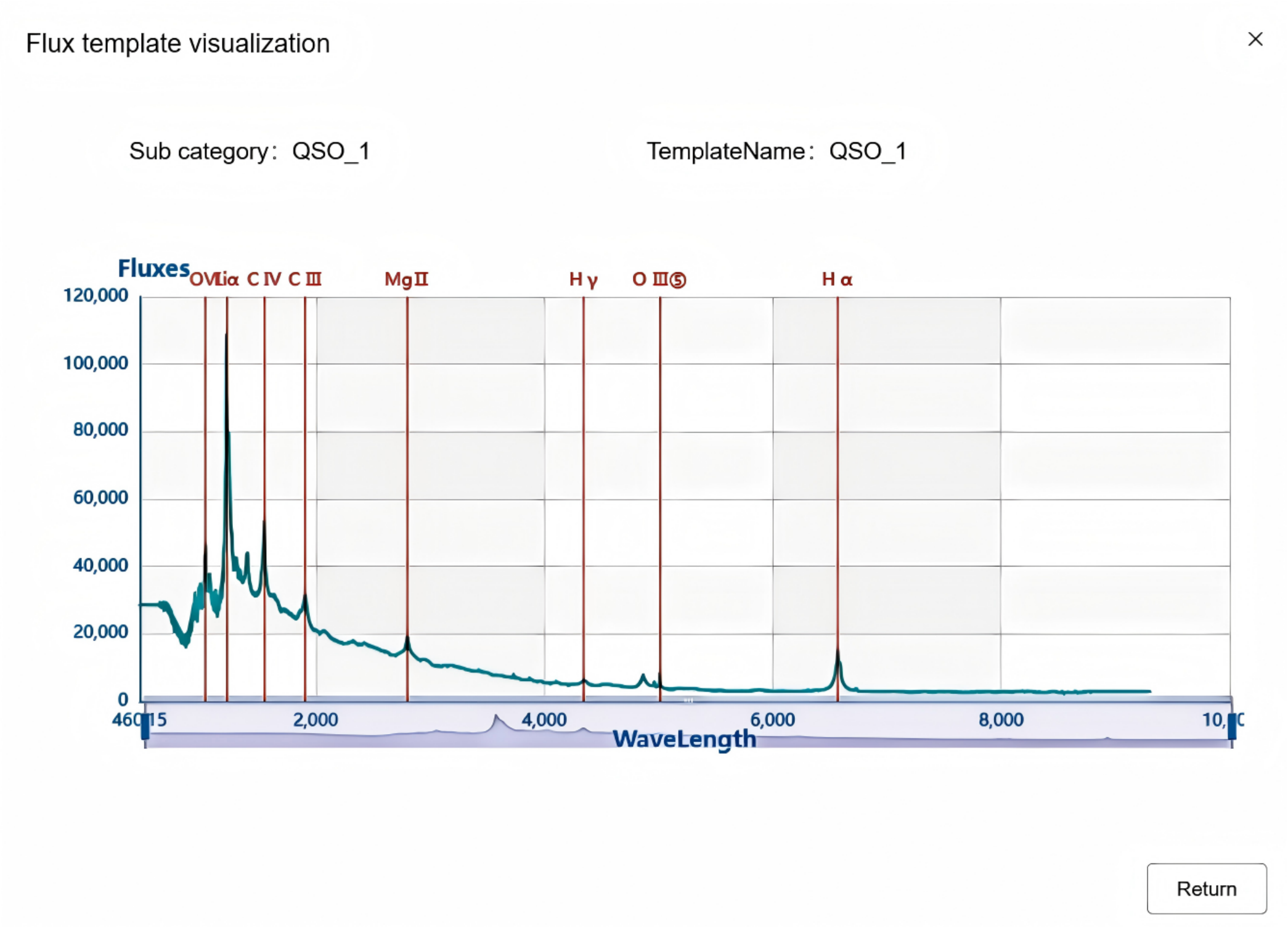}
\caption{{Spectral} %MDPI: 1. Please check if colors need extra explanation. 2. Please check if overlapped content doesn't affect scientific reading. 3. Please remove comma from numbers with 4 digits. 4. Please check if redundant content in the figure can be removed.(Thank you for the suggestions. This figure is a screenshot of the actual implemented system and serves only for interface demonstration. The colors and overlapping elements are part of the real interface and do not affect scientific reading. The commas in four-digit numbers and other elements are system defaults, and modifying them would require code changes. Since the figure is purely illustrative, we believe it is acceptable as is. Please let us know if any changes are strictly required.)
 template of a quasar from the Sloan Digital Sky Survey (SDSS).\label{fig:template}}
\end{figure}

{{Catalogue Data Management and Visualization.} 
 Users can upload catalogue data from LAMOST, SDSS, DESI, APOGEE~\citep{prieto2008apogee} and other sources in the standard CSV format. Uploaded data are automatically validated and parsed, enabling browsing, spectral analysis, and dataset comparison. SpecZoo automatically recognizes the unique identifiers of major sky surveys (e.g., OBSID for LAMOST; refer to the official survey manuals for identifier specifications of different projects) and downloads the corresponding spectral files via their URLs. Furthermore, the platform supports an Extralinks feature: if the uploaded CSV file contains RA and DEC coordinate fields, the system automatically reads them, allowing users to click on a survey identifier to directly navigate to its dedicated data query interface. Advanced visualization features include spectral line annotation, local feature zoom, and wavelet-based noise reduction.

{{Spectral Task Configuration and Annotation.} 
 Within the same workgroup, group leaders can assign research or educational tasks, and users can label the physical parameters of selected spectral samples. Typical parameters include redshift, emission line width, and continuum flux for quasars; redshift, mean age, metallicity, and total mass for galaxies; radial velocity, surface gravity, and metallicity for stars. Figure~\ref{fig:label} presents the task release interface, which enables multiple observers to conduct repeated observations of the same spectrum. Parameters can be dynamically adjusted, allowing tasks to be tailored to celestial objects while maintaining annotation rigor. SpecZoo supports multiple rounds of annotation by different users on the same spectrum, as well as the assignment of multiple spectrum to a single annotator concurrently.

\begin{figure}[H]
\includegraphics[width=0.65\linewidth]{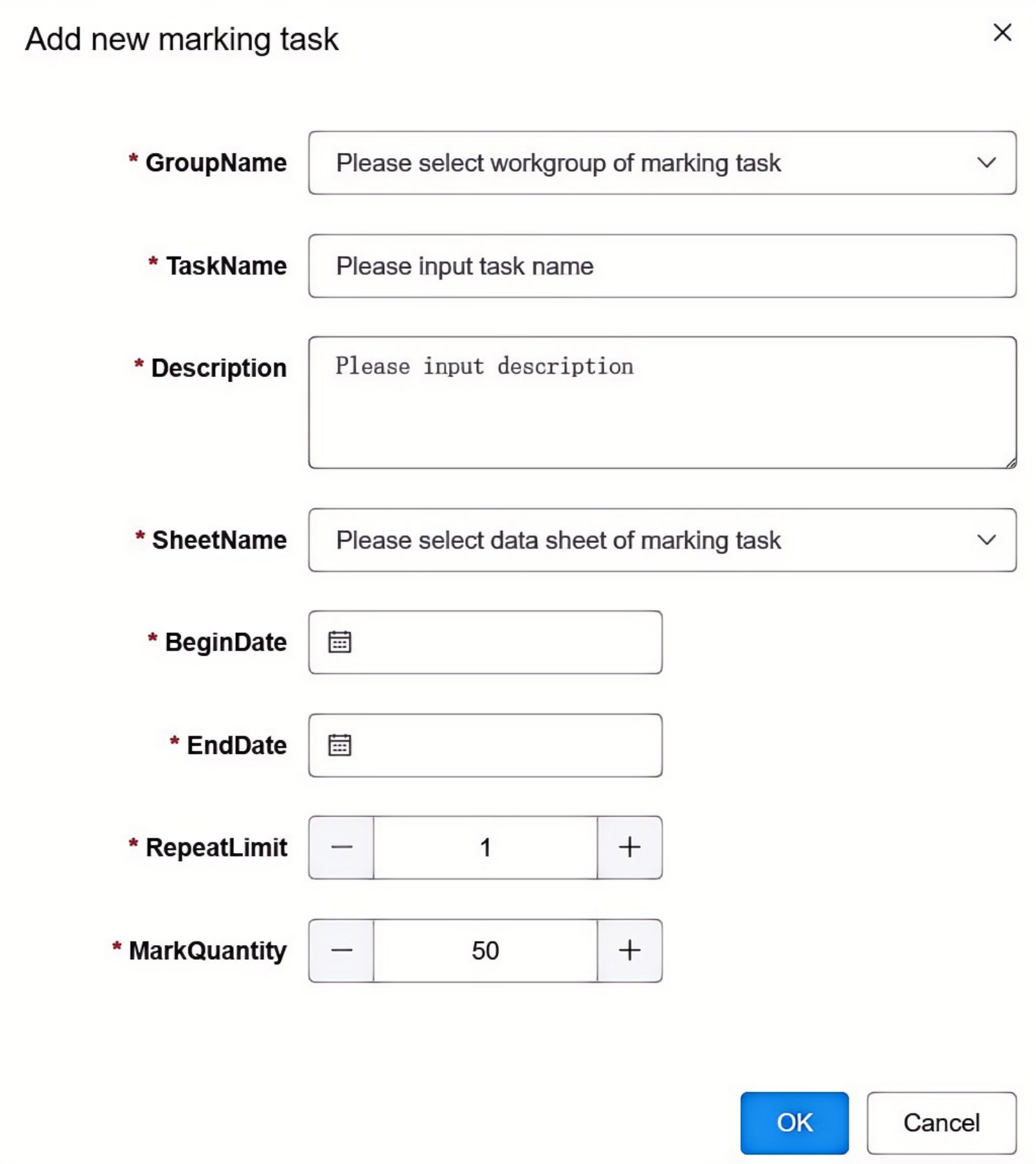}
\caption{{Interface} %MDPI: 1. Please check if redundant content in the figure can be removed. 2. Please check if asterisk need explanation. 3. Please confirm different font style. 4. We moved the figure after its first citation, please check and confirm.(check)
 for spectral recognition task labeling{(Asterisk indicates required field)}, enabling users to annotate physical parameters of selected spectral samples.\label{fig:label}}
\end{figure}

\subsection{Advanced AI-Powered Features}\label{3.2}

SpecZoo integrates multiple machine learning and deep learning frameworks to support automated spectral classification and parameter estimation.

{AI-powered Spectral Classification. } MSPC-Net~\cite{wu2024classification} is a convolutional neural network designed for automated spectral classification. It employs {multi-scale feature extraction} %MDPI: Please confirm if the italics is unnecessary and can be removed. Same for similar cases below.
 to capture both fine-grained local features and overall spectral structure, partial convolution to reduce overfitting by operating on channel subsets, and grouped convolution to enhance computational efficiency. This combination allows MSPC-Net to robustly classify large spectral datasets with consistently high accuracy {(e.g., 87--91\% on stellar classification and up to 85\% on subclass tasks)} and strong generalization, handling diverse object types including stars, galaxies, and quasars.

{Stellar Atmospheric Parameter Prediction.} SLAM (Stellar Label Machine)~\cite{zhang2020deriving} leverages Support Vector Regression to extract fundamental stellar parameters from spectral surveys. It dynamically adapts model complexity to different spectral types and signal-to-noise levels, making it suitable for both large-sample surveys and small-sample, resource-constrained studies. On the LAMOST DR5 dataset, SLAM achieves accuracies of 50\,K for $T_{\rm eff}$, 0.09\,dex for $\log g$, and 0.07\,dex for [Fe/H] on high S/N spectra. The method has been successfully applied to combined LAMOST-APOGEE data, demonstrating scalability and reliability across heterogeneous datasets.

{Integration with Large Language Models.} SpecCLIP~\cite{zhao2025specclip}, developed jointly by Zhejiang Lab and NAOC, combines natural language processing techniques with spectral analysis, enabling large-scale feature extraction and automated parameter estimation. SpecCLIP can process complex spectral patterns and low signal-to-noise data, complementing SLAM in a hybrid framework. This integration supports end-to-end analyses, anomaly detection, and rapid processing of extensive survey datasets. {SpecCLIP achieves accuracies of 132.669\,K for $T_{\rm eff}$, 0.079\,dex for $\log g$, and 0.056\,dex for [Fe/H] on LAMOST low‑resolution spectra.}

{GaSNet-III Spectral Analysis.} 
 GaSNet-III~\cite{zhong2024galaxy} is a generative deep learning model for spectral reconstruction, redshift estimation, and anomaly detection. Its architecture combines an autoencoder variant for interpretable feature extraction (analogous to PCA) with a U-Net structure for noise reduction, achieving high-fidelity reconstruction in logarithmic wavelength space. The model explores the $\chi^2$ space efficiently to identify global minima across object types and redshifts. For stellar spectra, classification accuracy exceeds 99.9\%, with overall accuracy above 98\%. Compared with traditional template-matching methods, GaSNet-III significantly improves computational efficiency and scalability for upcoming large surveys such as 4MOST~\cite{de20194most}.

The output parameters of the AI-powered functionalities in SpecZoo are specified in Table~\ref{Para}. {The symbol $\sigma$ represents the error rate. For the specific calculation method, please refer to the detailed literature on the three AI functions}~\cite{zhao2025specclip,zhang2020deriving}.

\begin{table}[H]
\small
\caption{{Physical} %MDPI: We moved the table after its first citation, please check and confirm.(check)
 and chemical parameters derived by SpecZoo. The table lists key stellar properties and asteroseismic quantities output by the platform.\label{Para}}
\begin{tabularx}{\textwidth}{CcCcC}
\toprule
\textbf{} & \textbf{Parameter} & \textbf{Unit} & \textbf{Description} & \boldmath{\textbf{$\sigma$}} \\
\midrule
\multirow{18}{*}{SpecCLIP} & Mass & $M_\odot$ & Stellar mass & 0.086 \\
 & Age & Gyr & Stellar age & 1.337 \\
 & rv & km\,s$^{-1}$ & Radial velocity & 5.289 \\
 & Teff & K & Effective temperature & 132.669 \\
 & $[\mathrm{Fe}/\mathrm{H}]$ & dex & Metallicity & 0.056 \\
 & $[\mathrm{C}/\mathrm{Fe}]$ & dex & Abundance ratio of C to Fe & 0.037 \\
 & $[\mathrm{O}/\mathrm{Fe}]$ & dex & Abundance ratio of O to Fe & 0.049 \\
 & $[\alpha/\mathrm{Fe}]$ & dex & Abundance ratio of $\alpha$ to Fe & 0.020 \\
 & $[\mathrm{N}/\mathrm{Fe}]$ & dex & Abundance ratio of N to Fe & 0.049 \\
 & $[\mathrm{Al}/\mathrm{Fe}]$ & dex & Abundance ratio of Al to Fe & 0.046 \\
 & $[\mathrm{Ca}/\mathrm{Fe}]$ & dex & Abundance ratio of Ca to Fe & 0.029 \\
 & $[\mathrm{Mg}/\mathrm{Fe}]$ & dex & Abundance ratio of Mg to Fe & 0.031 \\
 & $[\mathrm{Si}/\mathrm{Fe}]$ & dex & Abundance ratio of Si to Fe & 0.028 \\
 & $[\mathrm{Ti}/\mathrm{Fe}]$ & dex & Abundance ratio of Ti to Fe & 0.056 \\
 & $[\mathrm{Mn}/\mathrm{Fe}]$ & dex & Abundance ratio of Mn to Fe & 0.031 \\
 & $[\mathrm{Ni}/\mathrm{Fe}]$ & dex & Abundance ratio of Ni to Fe & 0.025 \\
 & $[\mathrm{Cr}/\mathrm{Fe}]$ & dex & Abundance ratio of Cr to Fe & 0.074 \\
 & log $g$ & dex & Surface gravity & 0.079 \\
\midrule
\multirow{4}{*}{SLAM} & Teff & K & Effective temperature & 50 \\
 & log $g$ & dex & Surface gravity & 0.09 \\
 & $\mathrm{[Fe/H]}$ & dex & Iron-to-hydrogen abundance ratio & 0.07 \\
\midrule
\multirow{4}{*}{GasNet-III} & Best-Fit\_CLASS & - & Matched astronomical classification & - \\
 & Best-Fit\_Z & - & Estimated redshift & - \\
 & Degeneracy & - & Robustness Parameter & - \\
 & min\_chi\_square & - & Goodness of fit & - \\
\bottomrule
\end{tabularx}
\end{table}

\subsection{An Example for Displaying the Main Interface and Validation of the AI Functionalities}

Based on the LAMOST survey for target selection, Li et al.~\citep{li2022four} performed a homogeneous elemental abundance analysis of 385 very metal-poor stars using high-resolution spectra obtained with the Subaru Telescope~\citep{kaifu1998subaru} and incorporating Gaia parallax data, resulting in a target catalog. Here, we select a source (LAMOST obsid: 203105173) as an example for demonstration to showcase the interface and functionalities of SpecZoo, as demonstrated in {Figure}~\ref{fig:interface}. The core interface of SpecZoo is organized into three main tabs. Upon entering the Visualization tab, users access Visualization functions for spectral display, denoising, and template fitting. The interface also features four dedicated AI buttons (as detailed in Section~\ref{3.2}) that enable automated classification, parameter estimation, and feature extraction, alongside other fundamental tools for interactive spectral inspection; the Data Browser tab presents the dataset from uploaded CSV files in tabular form and incorporates filtering and bookmarking capabilities to facilitate rapid data retrieval, where selected entries can be directly visualized in the corresponding spectral interface; the Superposition tab is dedicated to spectral superimposition, enabling comparative analysis of multiple spectra from repeated observations of a single source as well as spectral comparisons across similar astrophysical objects.

Table \ref{tab11} specified the values of stellar parameters obtained using the AI functionalities, which  closely approximate the values obtained from the high-resolution spectra provided by Li et al.~\citep{li2022four}.

\begin{figure}[H]
\includegraphics[width=\linewidth]{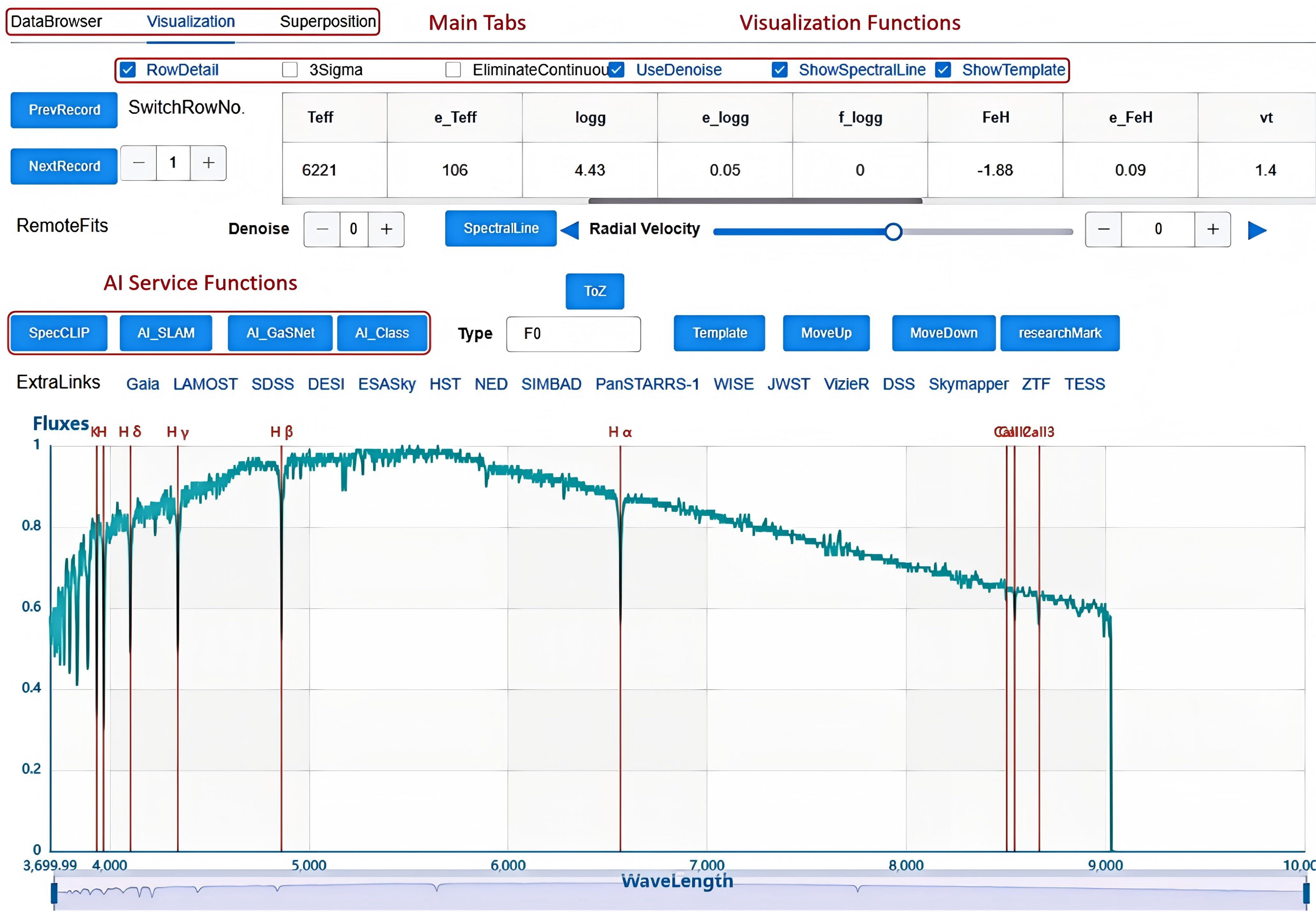}
\caption{{Core} %MDPI: 1. We revised the position of the figure in order to minimize the blank space left on the pages. 2. Please change the hyphen (-) into a minus sign ($-$, "U+2212"). e.g., "-1" should be "$-$1". 3. Please check if colors or symbols need explanation. 4. Please check if cropped content affect scientific reading. 5. Please remove comma from numbers with 4 digits.(check，I'm sorry, the commas in four-digit numbers come from the system interface, and modifying them would take a long time.)
 interface of SpecZoo illustrated with a metal-poor star. The interface provides three main tabs: `Data Browser' for navigating through all rows in the dataset; `Visualization' for detailed spectral analysis; and `Superposition' for comparing multiple spectra simultaneously. The four dedicated `AI buttons' enable the execution of distinct AI$-$powered functionalities, facilitating automated classification, parameter estimation, and feature extraction.\label{fig:interface}}
\end{figure}

\begin{table}[H]
%\small
\caption{A metal-poor star (LAMOST obsid: 203105173) is used as an example to demonstrate three algorithms. MSPC-Net shows it as F0 star, while SLAM and SpecCLIP measure its atmospheric parameters. \label{tab11}}
\begin{tabularx}{\textwidth}{cCCc}
\toprule
\textbf{Parameters} & \textbf{Values} &   \textbf{Literature Values} &      \textbf{Description} \\
\midrule
Teff & 5955.58 $\pm$ 215.07 & 6221 $\pm$ 106 & Effective temperature [K] \\
$[\mathrm{Fe}/\mathrm{H}]$ & $-$1.89 $\pm$ 0.23 & $-$.88 $\pm$ 0.09     & Metallicity [dex] \\
log $g$ & 4.14 & 4.43 $\pm$ 0.05     & Surface gravity [dex] \\
Best-Fit\_CLASS & STAR & STAR      & Matched astronomical classification \\
Best-Fit\_Z & $-$0.00011 & - & Estimated redshift \\
AI CLASS & F0 & - & AI recommends subcategories \\
\bottomrule
\end{tabularx}
\end{table}

%\section{Use Cases for Science} \label{sec:scientific}

\section{Use Cases for Science} \label{sec:scientific}
SpecZoo facilitates both independent and collaborative scientific investigations, particularly in the identification of rare objects and construction of statistically significant samples. The platform efficiently integrates multi-band spectral data, enabling rapid analysis and visualization for large datasets. For example, it has been demonstrated that the average time for graduate students to visually inspect a single spectrum can be reduced by $\sim$30$\%$ {(based on the team’s calculation of average screening time for special objects such as quasars and carbon stars using SpecZoo)}, allowing researchers to focus more on scientific interpretation and parameter extraction.

\subsection{Identification of Strong Gravitational Lens Candidates}

Strong gravitational lensing, a direct prediction of general relativity, occurs when a massive foreground object, such as a galaxy or black hole, bends the light from a more distant source, producing magnification, distortion, or multiple images. While most galaxy-scale lenses are traditionally identified through high-resolution imaging, automated detection using spectroscopic data remains a developing research frontier.

SpecZoo facilitates the efficient identification of strong gravitational lens candidates in large spectroscopic surveys by integrating spectral analysis, interactive visualization, and cross-survey data access. Within the platform, researchers can examine spectra to detect multiple redshift components and identify characteristic emission lines from background galaxies (e.g., [O~\textsc{iii}]~$\lambda5007$, Ly$\alpha$). Its cross-survey linkage module enables rapid retrieval of high-resolution imaging data {(e.g., Hubble Space Telescope (HST) deep-field observations}~\cite{brammer20123d}), allowing confirmation of lensing morphology such as Einstein rings, arcs, or multiple images.

As an illustrative example of manual visual screening strategies,~\cite{zhong2022galaxy} proposed a four-step workflow for large spectroscopic surveys: (1) assigning lens-candidate probabilities to all spectra; (2) selecting high-probability candidates above a predefined threshold; (3)~estimating redshifts for both foreground and background objects, requiring the background redshift to exceed the foreground by at least $0.1$; and (4) conducting manual visual inspection to verify candidate authenticity. While SpecZoo does not directly implement this specific framework, its combination of interactive spectral tools and candidate management capabilities supports analogous workflows, enabling researchers to efficiently perform manual validation of strong lens candidates. Specifically, SpecZoo facilitates rapid processing during the visual inspection phase; after uploading a catalog of candidates, researchers can leverage its interactive visualization tools to compare multiple spectra and annotations simultaneously, improving inspection efficiency several-fold.

Figure \ref{lens1} shows a representative strong gravitational lensing system identified using SpecZoo, accompanied by high-resolution HST imaging (insets). The foreground galaxy is a quiescent red galaxy exhibiting prominent absorption features, including K, H, Mg~I, Na~I, and H$\alpha$, with a measured redshift of $z = 0.1196$. Superimposed on this spectrum are emission lines from a background galaxy (Figure \ref{lens2}) at $z = 0.1965$, confirming the coexistence of foreground and background components and demonstrating the characteristic double-redshift signature of a strong lensing system.

\begin{figure}[H]
\includegraphics[width=1\linewidth]{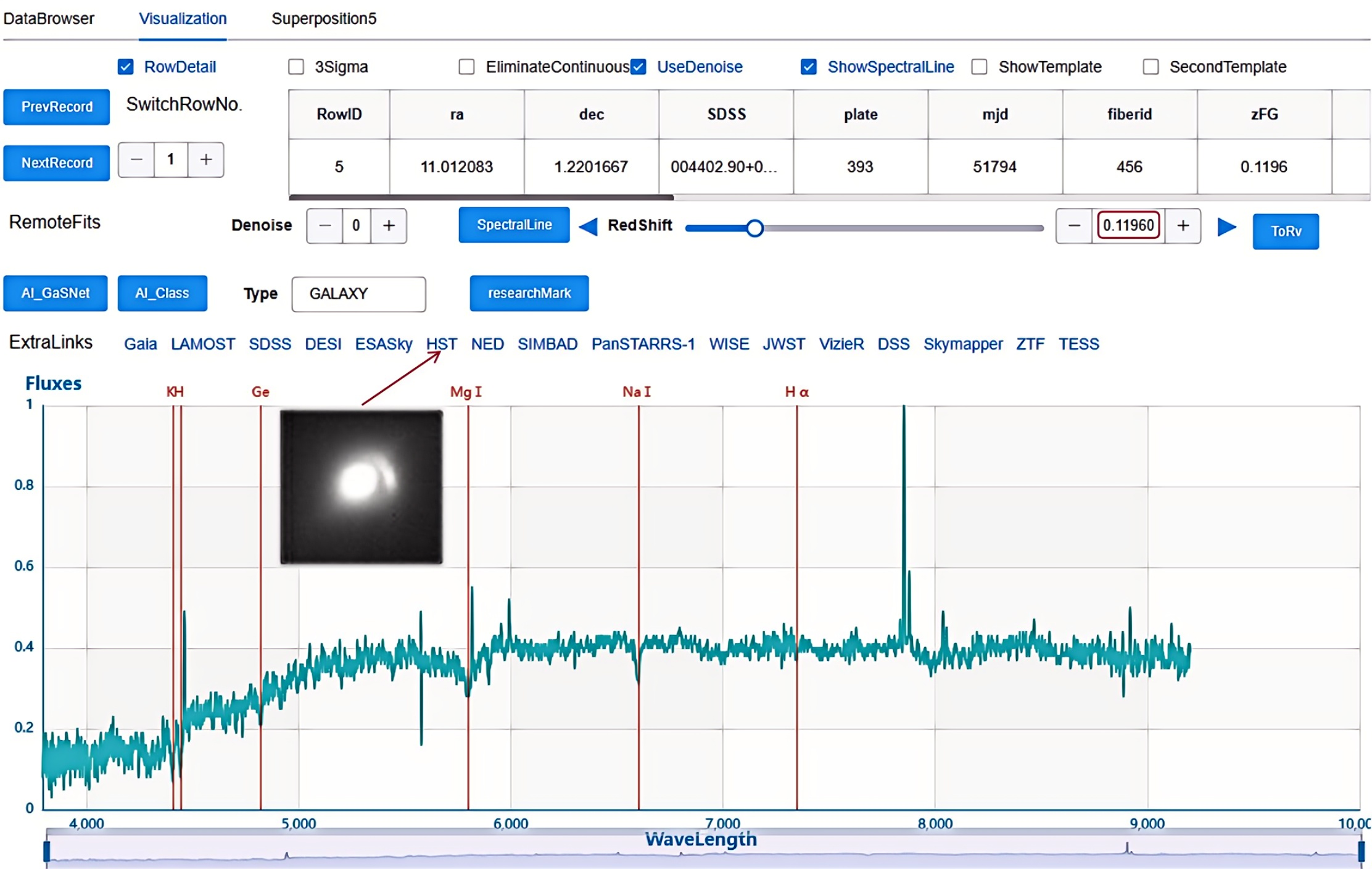}
\caption{{Spectra} %MDPI: 1. We moved the figure after its first citation, please check and confirm. same for figure 8. 2. Please check if colors or symbols need explanation. 3. Please check if cropped content affect scientific reading. 4. Please remove comma from numbers with 4 digits.(check,I'm sorry, the commas in four-digit numbers come from the system interface, and modifying them would take a long time.)
 {(cyan curve)}and HST images (insets) of a source exhibiting strong gravitational lensing. The foreground galaxy is an old red galaxy, with absorption lines of K, H, Ge, Mg I, Na I, and H$\alpha$, corresponding to a redshift of $z=0.1196$.\label{lens1}}
\end{figure}
\unskip
\begin{figure}[H]
\includegraphics[width=1\linewidth]{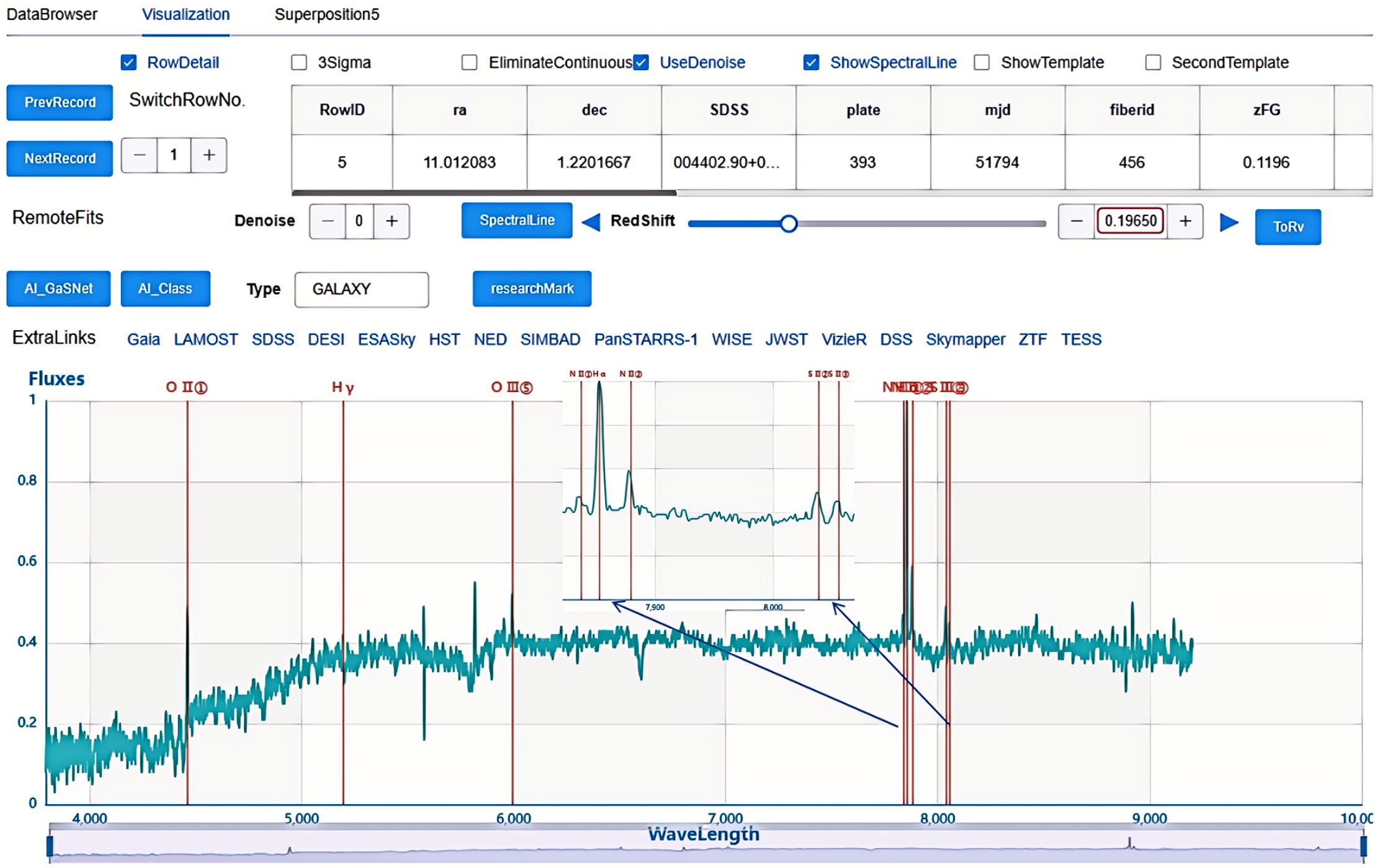}
\caption{{Emission} %MDPI: 1. Please check if colors or symbols need explanation. 2. Please check if cropped and overlapped content affect scientific reading. 3. Please remove comma from numbers with 4 digits.(check,I'm sorry, the commas in four-digit numbers come from the system interface, and modifying them would take a long time.)
 lines from the background galaxy overlay the spectrum of the strongly lensed source (red curve). The inset shows a magnified view around 8000\,\AA, highlighting the emission features corresponding to a redshift of $z=0.1965$.\label{lens2}}
\end{figure}

\textls[-15]{This example illustrates how SpecZoo facilitates both the identification and preliminary validation of strong lens candidates. The platform supports: (1) interactive detection of multiple redshift components within a single spectrum, (2) integration with high-resolution imaging to verify morphological features such as arcs or Einstein rings, and (3) efficient candidate management that combines automated screening with manual verification.}

\textls[-15]{Our team is developing a machine‑learning pipeline on the SpecZoo platform to systematically identify strong gravitational‑lens candidates within DESI spectroscopic data, primarily by detecting multiple, spatially unresolved redshift components within single‑fiber spectra (Li et al., in preparation). This spectroscopic search provides a strategic precursor catalog for next‑generation imaging surveys, most notably the China Space Station Telescope (CSST)~\citep{miao2023cosmological}. The synergy between DESI spectroscopy and CSST high‑resolution imaging will enable definitive confirmation and the construction of a robust, multi‑wavelength lens sample, exemplifying a forward‑looking approach to maximizing the scientific return of complementary survey technologies.}

\subsection{White Dwarf–Main Sequence Binary Systems}

\textls[-15]{White dwarf–main-sequence (WDMS) binaries are a common class of compact binaries in the Milky Way, each consisting of a white dwarf (WD) and a low-mass main-sequence (MS) companion. These systems typically form from main-sequence binary evolution. In wide binaries, the two components evolve largely independently with little mass transfer. In close binaries, the more massive star may expand during its red giant or asymptotic giant branch phase, engulfing its companion and initiating a common-envelope (CE) phase. {Tides within the envelope reduce the orbital separation and remove energy and angular momentum, eventually leading to envelope ejection and leaving behind a post-common-envelope binary (PCEB)}. Owing to their abundance and relative ease of identification, WDMS binaries serve as key testbeds for studying common-envelope evolution~\cite{Ren2014-WDMS-Review}.}

\textls[-15]{SpecZoo facilitates the identification and analysis of WDMS systems by combining interactive spectral analysis, template fitting, and cross-survey data integration. The platform's two-template fitting module allows decomposition of composite spectra, assigning separate spectral templates to distinct wavelength ranges. This enables disentangling of overlapping WD and main-sequence features, even when the angular separation is too small for fiber spectroscopy to resolve the components individually. Cross-matched astrometric data, including Gaia parallaxes and proper motions, can be retrieved within SpecZoo to verify spatial and kinematic consistency, providing an additional layer of confirmation for physically bound systems. Figure~\ref{fig:9} illustrates a representative WDMS system identified using SpecZoo. Panel shows the SDSS optical image, with the {cyan curve} representing the composite LAMOST spectrum. The WD component is matched to the blue portion of the spectrum (purple curve), while the M-type main-sequence companion is matched to the red portion (blue curve), demonstrating effective spectral disentanglement. Gaia parallaxes ($\pi_1=7.33\pm 0.07\,\mathrm{mas}$, $\pi_2=7.36\pm 0.11\, \mathrm{mas}$) and proper motions ($\mu_1=-20.52\pm 0.07\,\mathrm{mas \, yr^{-1}}$, $\mu_2=-20.44\pm 0.10\,\mathrm{mas \, yr^{-1}}$) for both components, confirming their spatial and kinematic coherence and validating the physical association.}

\begin{figure}[H]
\includegraphics[width=0.9\linewidth]{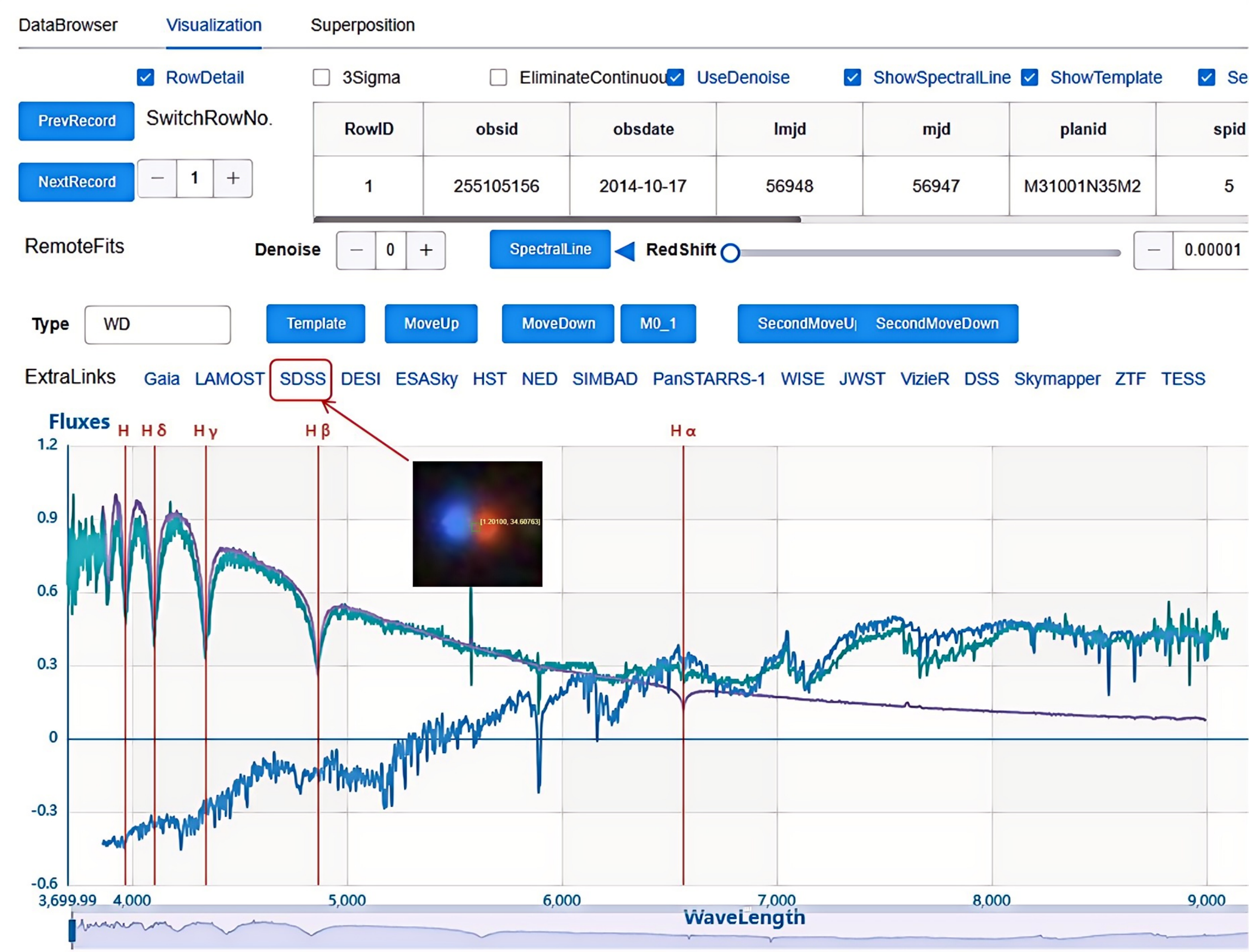}
\caption{{Analysis} %MDPI: 1. Please check if colors or symbols need explanation. 2. Please check if cropped content affect scientific reading. 3. Please remove comma from numbers with 4 digits. 4. Please change the hyphen (-) into a minus sign ($-$, "U+2212"). e.g., "-1" should be "$-$1". 5. Please check if elipsis should be revised.(check,I'm sorry, the commas in four-digit numbers come from the system interface, and modifying them would take a long time.)
 of a WDMS binary using SpecZoo. \textbf{Panel} LAMOST composite spectrum {(cyan curve)} overlaid with SDSS optical images; the purple and blue curves represent the white dwarf and M$-$type main$-$sequence templates, respectively, fitted via SpecZoo's two$-$template decomposition. Gaia parallaxes and proper motions for both components, enabling verification of their physical association.\label{fig:9}}
\end{figure}

\textls[-15]{This approach can be readily applied to other types of binary systems. By leveraging SpecZoo's capabilities for multimodal data integration and intelligent spectral analysis, researchers can efficiently identify and assemble large samples of binaries, facilitating statistical studies and evolutionary modeling. Systematic searches using spectra from SDSS, LAMOST, and other surveys are essential for investigating the evolution of compact binaries, particularly the physical processes governing the common envelope phase~\cite{willems2004detached}. SpecZoo's combination of spectral decomposition, template fitting, and cross-survey data access enables rapid identification, classification, and validation of rare or complex binary systems.}

During the visual inspection of WDMS binaries, our team identified a peculiar cataclysmic variable system that exhibits pronounced eclipsing behavior in its light curve, confirmed by combined spectral and time-domain analysis. We are currently modeling its orbital parameters and geometric structure {(Feng et al.,} %MDPI: Please check if a reference citation should be added.(not necessary,just in preparation)
 in preparation).

\subsection{Efficient Identification and Subtype Classification of Carbon Stars}

\textls[-15]{Carbon stars constitute a distinct population of late-type giants (spectral types G, K, and M) whose atmospheres are enriched in carbon relative to oxygen. This enhancement produces prominent molecular absorption features, including C$_2$, CN, and CH bands, which distinguish them from typical late-type giants of similar temperature and luminosity. Carbon stars play a critical role as tracers of stellar evolution and Galactic chemical enrichment. They are further subclassified into spectral types such as C--H, C--R, C--N, and Ba stars, reflecting variations in molecular absorption features, surface composition, and formation pathways, often linked to binary  mass-transfer events~\citep{1996ApJS..105..419B}.}

\textls[-15]{Traditionally, carbon star identification relies heavily on manual spectral inspection. For instance,~\cite{li2024identification} applied a multi-stage workflow to 10,599,979 low-resolution LAMOST DR7 spectra, including (1) pre-selecting spectra with $i$-band signal-to-noise ratio \mbox{$\mathrm{S/N}(i) > 10$;} (2) screening candidates using molecular line indices; (3) incorporating 2MASS infrared photometry to separate ``warm'' and ``cold'' subgroups; and (4) performing final manual verification. While effective, this approach requires extensive manual effort, yielding over 100,000 candidate spectra that demand  labor-intensive inspection.}

\textls[-15]{SpecZoo significantly enhances both efficiency and reliability in the identification and subtype classification of carbon stars. Candidate spectra pre-selected via molecular line indices and infrared colors can be interactively analyzed within the platform. SpecZoo enables detailed visualization of diagnostic molecular absorption features, including the C$_2$ bands (4737\,\AA, 5165\,\AA, 5635\,\AA), CN bands (7065\,\AA, 7820\,\AA), and continuum characteristics (e.g., flux cutoffs at $\lambda < 4400$\,\AA\ in C--N stars). Subtype classification is further facilitated through semi-automated identification of key spectral line features following established criteria~\citep{1996ApJS..105..419B}. By integrating automated pre-selection, spectral decomposition, and interactive validation, SpecZoo drastically reduces manual workload and supports the rapid assembly of large, robust carbon \linebreak  star samples.}

\textls[-15]{Figure~\ref{carbon} illustrates a representative carbon star candidate identified using SpecZoo. The spectrum, obtained from LAMOST survey data, exhibits strong C$_2$ absorption bands at 4737\,\AA\ and 5165\,\AA, along with pronounced s-process element lines, including Sr~II~4077\,\AA\ and Ba~II~4554\,\AA\ and 6496\,\AA, indicating a Ba-type carbon star. Interactive anaslysis with SpecCLIP quantifies the carbon-to-oxygen ratio, confirming that {the super-solar abundance of C/O.} This example demonstrates how, following automated pre-selection by external pipelines, SpecZoo combines spectral decomposition and quantitative assessment to efficiently process, identify, and classify carbon star candidates, thereby enabling large-scale studies of their formation history and chemical \linebreak  enrichment patterns.}

\begin{figure}[H]
\includegraphics[width=0.9\linewidth]{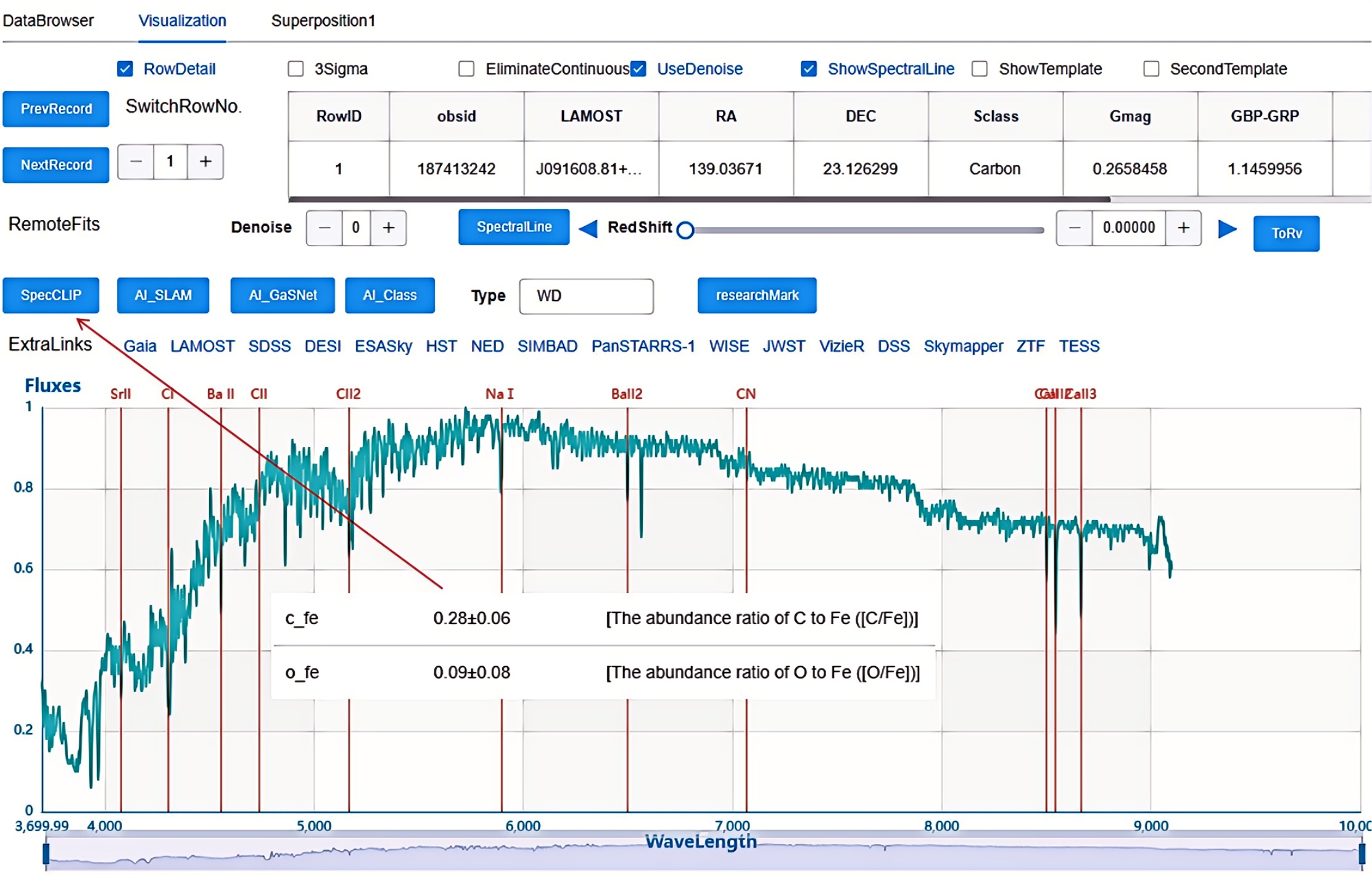}
\caption{{The spectrum} %MDPI: 1. Please check if colors or symbols need explanation. 2. Please check if cropped content affect scientific reading. 3. Please remove comma from numbers with 4 digits. (check,I'm sorry, the commas in four-digit numbers come from the system interface, and modifying them would take a long time.)
 of a carbon star candidate from LAMOST reveals strong C\,II molecular absorption bands at 4737\,\AA\ and 5165\,\AA\, along with strong absorption lines of Sr\,II (4077\,\AA\ ) and Ba\,II (4554\,\AA\ , 6496\,\AA\ ), identifying it as a Ba-type carbon star. According to the results returned by SpecCLIP (subplot), {the super-solar abundance of C/O}.\label{carbon}}
\end{figure}

\section{Use Cases for Education}\label{sec:education}

SpecZoo also has educational functions, which contributes to the cultivation of astronomical talents and astronomical education that proceeds from the elementary to the advanced.

\subsection{Assist in Foundational Teaching}

In the first year of undergraduate course \emph{Astronomical Data Processing} at China West Normal University, SpecZoo was fully integrated into the module on stellar spectral classification. The platform supports a structured workflow in which students first employ the MSPC-Net AI classification module to predict the spectral types of unknown stars from large survey data, such as LAMOST or SDSS (e.g., F, G, K, and M-type stars). These preliminary AI-powered predictions provide efficient and reliable guidance for subsequent analysis. Students then perform manual validation by comparing the target spectra with the platform's comprehensive library of spectral templates based on MKCLASS~\citep{jaschek1964catalogue} standards. Overlaying the spectra with templates allows detailed inspection of line morphology, intensity, and wavelength positions of diagnostic absorption features (e.g., Balmer series, metallic lines), facilitating a quantitative understanding of stellar physical properties such as effective temperature and surface gravity.

The integration of AI significantly accelerates the learning process. In practice, students achieved an average classification accuracy of {69\%} after only 16 hours of training ({see Figure}~\ref{fig:placeholder}), compared to approximately 40\% for traditional astronomical pedagogy. This dual approach---{AI-assisted pre-classification followed by manual template verification}%MDPI: Please confirm if the italics is unnecessary and can be removed. The following highlights are the same.
---lowers the entry barrier for spectral analysis and reduces the time required compared with traditional fully manual classification. Importantly, it transforms abstract theoretical concepts into hands-on practice, enabling students to systematically develop spectral diagnostic skills and deepen their understanding of the connection between stellar physical states and observed spectral features. With continued practice, students are able to rapidly classify stellar spectra by recognizing key absorption features.

\begin{figure}[H]
\includegraphics[width=0.9\linewidth]{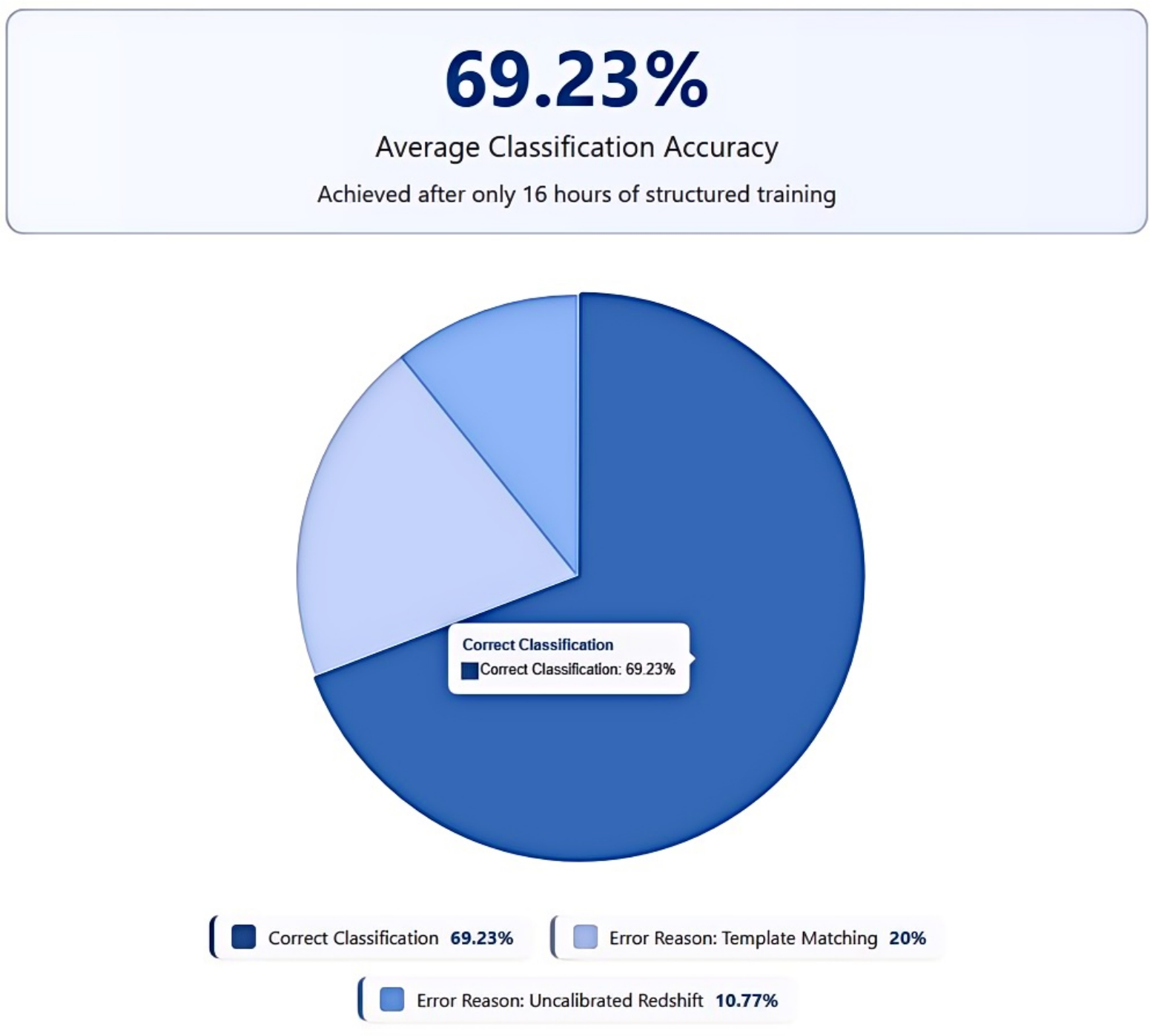}
\caption{Undergraduate course ``Astronomical Data Processing'' teaching effectiveness visualization based on 16 hours of structured AI-assisted training.\label{fig:placeholder}}
\end{figure}

\subsection{Support for Research-Oriented Teaching}

In the {Research Training and Innovative Practice } course launched for the Physics Elite Class\endnote{The class, which enrolls around 30 junior physics majors each year, prepares students for careers in scientific research.} at Hangzhou Dianzi University, the curriculum is designed around the analysis of massive spectroscopic datasets assisted by artificial intelligence, guiding students in the systematic search for rare celestial objects (e.g., Be stars~\citep{slettebak1979stars}, WDMS binaries~\citep{Ren2014-WDMS-Review}, carbon stars~\citep{li2024identification}, strong gravitational lens~\citep{zhong2022galaxy}, and so on). The course utilizes the SpecZoo expert spectroscopic platform to conduct hands-on training in spectral visualization and analysis. {Students are divided into 4--6 research groups (with 4--5 students per group)}, each tasked with replicating the complete search pipeline for distinct types of celestial objects, such as carbon stars, quasars, and WDMS binaries, and so on. This process covers the full research cycle, including literature review, data processing, model construction, and result validation. Through an integrated teaching model combining theory, practice, and discussion, the course aims to cultivate foundational research competencies among students. Promising projects are further encouraged to extend into undergraduate thesis work or graduate research, fostering a continuous talent development pipeline.

Based on pedagogical assessments conducted during the 2024 and 2025 academic years, over 95\% of students entered the course with no prior background in spectral analysis. Following a structured 64 h training program supported by the integrated tools of the SpecZoo platform, all student groups achieved the ability to independently conduct spectral analyses and fully reproduce the object‑discovery or special object validation workflows documented in the reference literature. Notably, several groups extended their work to produce original research outcomes, for {instance,} %MDPI: Please check if comma should be revised with colon.(check)

\begin{enumerate}
\item[(1)] {The} %MDPI: We revised this as list format, please check and confirm.(check)
 gravitational‑lens group performed a systematic search of 220,000 LAMOST galaxy spectra with the machine‑learning approach, identifying 170 candidate strong lenses. Subsequent manual verification on SpecZoo yielded around 20 high‑probability lens candidates.
\item[(2)] The carbon‑star group manually reviewed and subclassified existing carbon‑star catalogs via SpecZoo, optimized the deep‑learning algorithm from~\citet{he2024identification}, and applied it to the latest LAMOST data release. 
\item[(3)] The WDMS group reproduced the methodology of~\citet{perez2025finding} and successfully adapted its unsupervised machine‑learning approach to low‑resolution LAMOST spectra for WDMS candidate detection.
\end{enumerate}

The aforementioned findings and methodological innovations in each direction demonstrate notable academic merit and publication potential. The involved student teams are currently systematically organizing their research findings and actively preparing manuscripts for corresponding academic papers.

These results demonstrate that, through hands‑on use of the SpecZoo platform, students can effectively acquire and apply contemporary research methodologies to real astronomical data. All final projects have been compiled into course papers and are publicly accessible on the SpecZoo platform homepage, illustrating the platform’s role in facilitating research‑oriented education and enabling novice learners to transition into active contributors in spectral data analysis.

\section{Conclusions}
\label{sec:conclusion}
Spectroscopic observations provide fundamental insights into the physical properties and chemical compositions of astronomical objects, serving as a cornerstone for studies of stellar populations, galactic structure, and cosmic evolution. We developed an integrated spectral analysis platform, SpecZoo, that combines artificial intelligence, interactive visualization, and multimodal data integration to support both scientific research and education. The platform offers automated spectral classification, precise stellar parameter estimation, spectral decomposition, cross-survey data access, interactive visualization, and manual annotation, enabling efficient identification of astrophysical objects, construction of large statistical samples, and systematic analyses.

The scientific capabilities of SpecZoo are illustrated through three representative applications. First, in the identification of strong gravitational lens candidates, the platform enables spectroscopic detection of multiple redshift components within a single observation, allowing researchers to disentangle foreground and background sources and efficiently screen and validate candidate systems. While high-resolution imaging can provide morphological confirmation, the primary identification relies on the characteristic spectral redshift signatures. Second, for WDMS binaries, SpecZoo facilitates spectral decomposition and template fitting to separate overlapping signals, which, when combined with parallax and proper motion measurements, confirms physically bound systems. This enables the assembly of large, statistically robust samples for studies of binary evolution and the common envelope phase. Third, in the classification of carbon stars, the platform supports interactive visualization of molecular absorption features and s-process element lines, automated subtype identification, and cross-survey confirmation, greatly accelerating the construction of large, reliable samples for investigations of stellar evolution and galactic chemical enrichment. Each of these examples demonstrates how SpecZoo integrates automated analysis, human expertise, and multi-survey data to streamline discovery and validation workflows.

In educational contexts, SpecZoo converts theoretical concepts into hands-on practice. Through foundational teaching, students can quickly get started and significantly improve their spectral classification accuracy; through research-oriented teaching, students can rapidly become proficient and engage in actual scientific research tasks. Specifically, students use AI-assisted pre-classification (e.g., via MSPC-Net) to obtain preliminary spectral types, followed by interactive template verification and manual inspection, comparing spectral morphology, line strengths, and wavelength positions. This workflow not only lowers the barrier to spectral analysis but also systematically develops students’ skills in spectral diagnostics, interpretation of stellar physical parameters, and scientific reasoning. With this dual AI–human approach, students gain both operational experience and enhanced understanding of the relationships between stellar spectra and underlying astrophysical properties.

By supporting collaborative annotation and feature identification, SpecZoo also generates high-quality labeled datasets that improve automated classification accuracy and enable iterative refinement of AI algorithms. Building upon this foundation, future work will leverage machine-learning techniques and the SpecZoo platform to systematically search for rare celestial objects. Future developments will incorporate advanced neural network architectures and generalized modeling frameworks to enhance multi-band and multimodal analysis. Expanded collaborations with institutions (e.g., NADC and Zhejiang Laboratory) will enable seamless cross-platform integration, broaden accessibility, and facilitate the construction of large, robust datasets. This research activity will be deeply integrated into research-oriented teaching, with the ultimate goal of establishing SpecZoo as a comprehensive spectral “zoo” for both scientific and educational communities. Overall, SpecZoo exemplifies the potential of AI-powered spectroscopy to advance astronomical research, support systematic studies of rare and complex objects, and cultivate the next generation of researchers through immersive, data-intensive learning experiences.

\vspace{6pt}

\authorcontributions{Conceptualization, H.-J.T.; methodology, Y.-H.P. and G.-H.L.; software, \mbox{G.-H.L.} and Y.X.; validation, Y.-H.P., G.-H.L. and H.-J.T.; formal analysis, H.-J.T. and Y.X.; investigation, Y.-H.P., Y.X. and H.-J.T.; resources, Y.-H.P. and H.-J.T.; data curation, Y.X. and Y.-H.P.; writing---original draft preparation, Y.-H.P.; writing---review and editing, X.-Z.C., Y.-H.P. and H.-J.T.; visualization, G.-H.L.; supervision, H.-J.T.; project administration, H.-J.T. and Y.-H.P.; funding acquisition, H.-J.T. and X.-Z.C. All authors have read and agreed to the published version of the manuscript.All authors have read and agreed to the published version of the manuscript.}

\funding{This work was supported by the National Natural Science Foundation of China (under Grant Nos. 12373033, 12447204, 12403037) and the first batch of scientific research projects of the China Space Station Telescope (CSST) (under Grant Nos. CMS-CSST-2021-A09, A08).}

\dataavailability{The data presented in this study are available upon reasonable request from the corresponding author. The spectral data analyzed in this work can be accessed through the SpecZoo system {on 1 January 2026 at} %MDPI: Please add the access date (format: Date Month Year), e.g., accessed on 1 January 2020.(check)
 \url{https://nadc.china-vo.org/speczoo-system} after registration and login. Due to the proprietary nature of the database and access restrictions, the raw data are not publicly available but can be obtained following the platform's data access protocol.}

\acknowledgments{{The authors thank Feng Wang, Jian-Rong Shi, Dong-Wei Fan, Chang-Hua Li, Yan-Xia Zhang, Chen-Zhou Cui and Hua-Xi Chen for the helpful discussions and thank Jing-Jing Wu, Fu-Cheng Zhong, and Han Wu for the deployment of the AI services. H.J.T. thanks the support from the Key Project of Zhejiang Provincial Natural Science Foundation (No. ZCLZ25A0301). Y.X. thanks the support by the Young Data Scientist Program of the National Astronomical Data Center (NADC)}.} %MDPI: We have detected similar duplicate content in the Acknowledgments and Funding sections. Please remove content from the Acknowledgments section that has already been covered in the Funding section.
}

\conflictsofinterest{The authors declare no conflicts of interest. The funders had no role in the design of the study; in the collection, analyses, or interpretation of data; in the writing of the manuscript; or in the decision to publish the results.}

%%%%%%%%%%%%%%%%%%%%%%%%%%%%%%%%%%%%%%%%%%
%\isPreprints{}{% This command is only used for ``preprints''.
\begin{adjustwidth}{-\extralength}{0cm}
\printendnotes[custom]
\reftitle{References}

% Please provide the correct journal abbreviation (e.g. according to the “List of Title Word Abbreviations” http://www.issn.org/services/online-services/access-to-the-ltwa/).

%\bibliography{ref}

\PublishersNote{}
%\isPreprints{}{% This command is only used for ``preprints''.
\end{adjustwidth}
\end{document}